\documentclass[twocolumn]{emulateapj} 

\newcommand{\pard}[2]{\frac{\partial #1}{\partial #2}}
\newcommand{\vek}[1]{\mbox{\boldmath$#1$}}

\begin{document}

\title{Spontaneous current-layer fragmentation and cascading
  reconnection in solar flares: I. Model and analysis}

\author{Miroslav B\'{a}rta\altaffilmark{1,2}, 
J\"{o}rg B\"{u}chner\altaffilmark{1}, 
Marian Karlick\'{y}\altaffilmark{2}, and
Jan Sk\'ala\altaffilmark{2,3}}
\affil{$^1$Max Planck Institute for Solar System Research,
D-37191 Katlenburg-Lindau, Germany\\
$^2$Astronomical Institute of the Academy of Sciences of the Czech
Republic, CZ-25165 Ond\v{r}ejov, Czech Republic\\
$^3$University of J.E. Purkinje, CZ-40096 \'{U}st\'{\i} nad Labem,
Czech Republic}
\email{barta@mps.mpg.de}


\begin{abstract}
Magnetic reconnection is commonly considered as a mechanism of solar
(eruptive) flares. A deeper study of this scenario reveals, however,
a number of open issues. Among them is the fundamental question, how the
magnetic energy is transferred from large, accumulation scales
to plasma scales where its actual dissipation takes place. 
In order to investigate this transfer over a broad range of scales we
address this question by means of a high-resolution MHD
simulation. The simulation results indicate that the magnetic-energy
transfer to small scales is realized via a cascade of
consecutive smaller and smaller flux-ropes (plasmoids), in analogy with the
vortex-tube cascade in (incompressible) fluid dynamics. Both tearing
and (driven) ``fragmenting coalescence'' 
processes are equally important 
for the consecutive fragmentation of the magnetic field 
(and associated current density) to
smaller elements. At the later stages a dynamic balance between tearing
and coalescence processes reveals a steady (power-law) scaling
typical for cascading processes. 
It is shown that cascading reconnection also addresses 
other open issues in solar flare research such as
the duality between the regular large-scale picture of (eruptive) flares and the
observed signatures of fragmented (chaotic) energy release, as well as the
huge number of accelerated particles. Indeed, spontaneous
current-layer fragmentation and formation of multiple channelised  
dissipative/acceleration regions embedded in the current layer
appears to be intrinsic to the cascading process. The multiple 
small-scale current sheets may also facilitate the acceleration of a 
large number of particles. The structure, distribution and dynamics of 
the embedded potential acceleration regions in a current layer fragmented
by cascading reconnection are studied and discussed.
\end{abstract}



\section{Introduction}
\label{sect:intro}

It is generally conjectured that solar flares represent a dissipative
part of the release of the magnetic energy 
accumulated in active regions at the Sun. 
The 'standard'  CSHKP model 
\citep[see, e.g.][and references therein]{Shibata+Tanuma:2001,Mandrini:2010, 
Magara+:1996} 
agrees well with the observed large-scale dynamics of eruptive events. 
In this model the flares are initiated by eruption of a flux-rope 
(in many cases observed as a filament)
via, for example, a kink or torus instability 
\citep[e.g.][]{Kliem+Toeroek:2006, Toeroek+Kliem:2005, Williams+:2005, 
Kliem+:2010} evolving later into a Coronal Mass Ejection (CME).
The latter is trailed by a large-scale current layer behind the ejecta 
\citep{Lin+Forbes:2000}. In this trailing current layer reconnection is 
supposed to give rise to various 
observed phenomena like hot SXR/EUV flare loops rooted in 
H$\alpha$ chromospheric ribbons, HXR sources in loop-top and 
foot-points, and radio emissions of various types. Some authors 
\citep[e.g.][]{Ko+:2003, Lin+:2007} argue that the
bright thin ray-like structure observed sometimes behind CMEs may represent
a manifestation of the density increase connected with this current sheet (CS).

However, closer analysis of the classical CSHKP model revealed some of its open
issues. Namely, the time-scales of reconnection in such a thick flare CS
appeared to be much longer than typical flare duration. In other words,
reconnection rate in such configurations has been 
found to be insufficient for the rapid
energy release observed in flares. Later it was found that   
the dissipation necessary for reconnection in the practically
collisionless solar corona is an essentially plasma-kinetic process 
\citep[see, e.g.,][]{Buchner:2006} 
which takes place at very small spatial scales. Hence, the question arises, 
how sufficiently thin CSs can build up within the 
global-scale, thick CME-trailing current layer: Open is the actual physical 
mechanism that provides the energy transfer from the global scales,
at which the energy is accumulated to the much smaller scales, at
which the plasma-kinetic dissipation takes place.

Addressing these questions \citet{Shibata+Tanuma:2001}
suggested a concept of cascading (or \textit{fractal}, as they call it)
reconnection. According to their scenario a 
cascade of non-linear tearing instabilities occurs in the continuously stretched
current layer formed behind a CME. 
Multiple magnetic islands
(helical flux-ropes in 3D), also called plasmoids, are formed, interleaved by
thin CSs. Due to increasing separation of the
plasmoids in the continuously vertically extending trailing part of
the CME the interleaving CSs are subjected to further filamentation until the
threshold for secondary tearing instability is reached. This process continues
further, third and higher levels of tearing instabilities take place,
until the width of the CSs reaches the kinetic scale.

This scenario has recently been supported by the analytical theory of
\textit{plasmoid instability} by \citet{Loureiro+:2007}. They show that the
high-Lundquist-number systems with high enough current-sheet
length-to-width ratio are not subjected to the slow Sweet-Parker
reconnection but they are inherently unstable to formation of
plasmoids on very short time-scales. \citet{Samtaney+:2009},
\citet{Bhattacharjee+:2009}, and \citet{Huang+Bhattacharjee:2010} confirmed
predictions of this analytical theory by numerical
simulations with high Lundquist numbers. \citet{Ni+:2010}
generalizes the model by presence of shear flows around current sheet (CS). 
\citet{Uzdensky+:2010} relate the theory of plasmoid instability
further to the concept of fractal reconnection suggested by 
\citet{Shibata+Tanuma:2001}. 
\citet{Shepherd+Cassak:2010} and \citet{Huang+:2010} study the
plasmoid instability numerically at smaller scales and investigate its 
relation to the Hall reconnection. They found various regimes of
parameters where different type of reconnection prevails.

Eventually, however, kinetic scales are reached where dissipation 
and particle acceleration take place most likely via kinetic
coalescence of micro-plasmoids and, possibly, their shrinkage 
\citep{Drake+:2005, Karlicky+Barta:2007, Karlicky+:2010}.

In addition to the issue of energy transport there are 
also other questions that remain open in the CSHKP model.
It is its apparent insufficiency to accelerate such a number of
particles in its single diffusion region
around the X-line that would correspond to the fluxes inferred
from HXR observations in the thick-target model 
\citep[][and references therein]{Fletcher:2005,Krucker+:2008}. 
And, furthermore, the HXR and radio
\citep[e.g. decimetric spikes, see][]{Karlicky+:1996,Karlicky+:2000,
Barta+Karlicky:2001} observations indicate that the particle acceleration
takes place via multiple concurrent small-scale events distributed chaotically
in the flare volume rather than by a single compact acceleration process hosted
by a single diffusion region. Such observations are usually referred to as 
signatures of fragmented/chaotic energy release in flares.

Because of these difficulties an alternative concept based on the so-called 
``self-organized criticality'' (SOC) has been proposed 
\citep{Aschwanden:2002, Vlahos:2007}. This class of models is based on the
idea of multiple small-scale CSs embedded in chaotic (braided)
magnetic fields that are formed as a consequence of random motions at the
system boundary (photosphere). Multiple CSs can host multiple reconnection
sites what provides natural explanation of observed signatures of fragmented
energy release. At the same moment they provide larger total volume of
diffusion regions, perhaps sufficient to account for the observed particle
fluxes also quantitatively. 
Organised large-scale picture (i.e. coherent structures like the flare-loop
arcades) should be 
in the case of SOC-based models achieved by the so-called avalanche principle: 
A small-scale energy release event can trigger similar events in its vicinity
provided the system is in marginally stable state (due to the continuous
pumping of energy and entropy through the boundary). Nevertheless, it is
difficult to achieve such 
coherent large-scale structures as they are
usually observed in solar flares in the frame of SOC-based models.

Thus, solar flares appear to be enigmatic phenomena exhibiting duality 
between the regular, well-organized dynamics of flares observed at large 
scales and signatures of fragmented/chaotic energy release seen in
observations related to flare-accelerated particles. 
While the coherent global eruption (flare) picture seems to be in agreement with
the CSHKP scenario, the observed
fragmented-energy-release signatures favor the SOC-based class of models.


\begin{figure}[t]
\epsscale{0.97}
\plotone{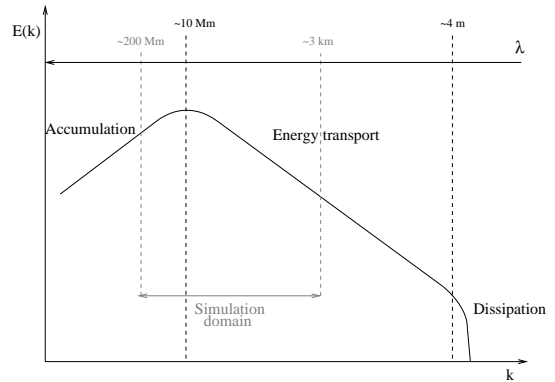}
\caption{Large-scale magnetic reconnection from the point of view of
  theory of dynamical systems. Schematic view of the
  cascade of energy transfer from large to small scales and the window
  of scales resolved in our simulations. Lower abscissa shows the wave number 
  $k$, corresponding characteristic scales $\lambda=2\pi/k$ are at the upper
  abscissa. The numeric values at the scale axis correspond
  to scaling based on typical coronal parameters (for details on
  scaling see Section~\ref{sect:model}).}
\label{fig:transfer}
\end{figure}


In the present paper, we suggest that cascading reconnection 
can address these three pressing questions (i.e. energy transport across
the scales, accelerated-particles fluxes, and the organized/chaotic picture
duality) 
as closely related to each other. 
In our view the energy is transferred from large to small scales by 
the cascade of fragmentation
of originally large-scale magnetic structures to smaller elements. 
We identify two
elementary processes of this fragmentation (see Section~\ref{sect:results}).
In the course of this process also the initial
current layer fragments into multiple small-scale, short-living current 
sheets.
These current sheets are hierarchically embedded inside the
thick current layer in qualitatively self-similar manner. In this
sense the cascading fragmentation reminds SOC models,
but now the chaotic distribution of small-scale currents results from
\textit{internal} instabilities of the global current layer. 
The fragmented current layer represents the modification of the standard CSHKP
model and thus it keeps coherent large-scale picture of solar flares.
At the same
time it addresses the observed signatures
of fragmented energy release and the question of efficient particle 
acceleration. We believe that the cascading reconnection in solar
coronal current layers can thus address the three main
problems mentioned above \textit{en bloc}, and it reconciles the two concepts
of the standard CSHKP and SOC-based models seen hitherto as antagonistic. 

The paper is organized as follows: First, we describe the model used in
our investigations. Then we present results of 
our high-resolution MHD simulation of cascading reconnection in an
extended, global, eruption-generated 
current layer. We identify the processes that lead to the fragmentation
of magnetic and current structures to smaller elements. 
Then we analyze the resulting scaling law of
the energy cascade. 
We describe the structure, distribution and
dynamics of small dissipation regions embedded in an initially thick current
layer. Finally, we discuss the implications that cascading reconnection have
for theory of solar flares.


\section{Model}
\label{sect:model}

Generally speaking, the solar flare involves three kinds of processes
that take place in different scale domains -- see Fig~\ref{fig:transfer}. 
At the largest scales magnetic-field energy is accumulated. 
During this stage flux-rope (filament) is formed and its magnetic energy
increases. Eventually it looses its stability and gets ejected. This
process already represents (ideal) release of the magnetic-energy
at large scales.
Subsequently, a current layer is formed and stretched
behind ejected flux-rope. Energy transfer from large scales at which the 
magnetic energy has
been accumulated to the small dissipation
scale occupies intermediate range of scales. The dissipation itself
takes place at smallest, kinetic scales.
 
In this paper we aim at studying energy transfer from large to small
scales by means of numerical simulations.
Despite the high spatial resolution our simulation is still 
within the MHD regime.
Also, we do not address the very process of energy
accumulation -- i.e. the flux-rope formation
and energization, nor its instability and
subsequent current-layer formation. Instead we assume a relatively
thick and extended current layer to be already formed at the initial
state of our study. 
In order to cover a large range of scales we limit ourselves
to the 2D geometry allowing for all three components of velocity and
magnetic field (commonly referred to as 2.5D models). This is a
reasonable assumption since observations show that the typical length of flare
arcades along the polarity inversion line (PIL) is much larger than
the dimension across the PIL.


\begin{figure}[t]
\epsscale{1.1}
\plotone{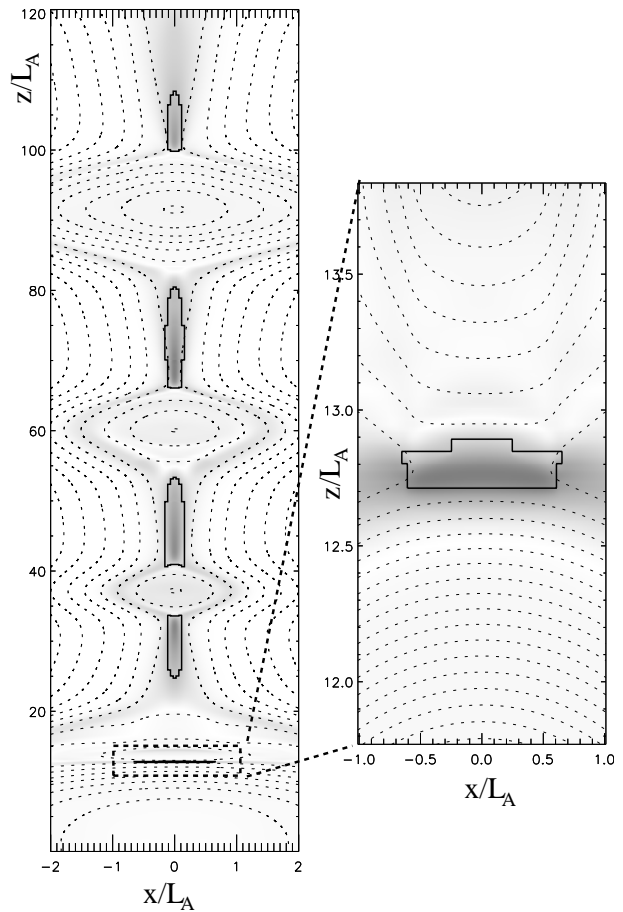}
\caption{Regions of adaptively enhanced resolution (thick-black-line
  bounded areas) on the background of current density magnitude
  (gray-scale) and magnetic field lines (dashed). The right panel
  shows zoomed view of the selected rectangle. Only the relevant
  sub-set of whole computational domain is shown, note the strongly
  anisotropic axes-scaling selected in order to show the high-resolution
  sub-domains better.}
\label{fig:zoom_cells}
\end{figure}


In the range of scales that we are interested in the evolution of
magnetized plasma can 
be adequately described by a set of compressible 
resistive one-fluid MHD equations \citep[e.g.][]{Priest:1984}:
\begin{eqnarray}
\nonumber
\pard{\rho}{t}+\vek{\nabla\cdot}(\rho\vek u)=0\\
\label{eq:MHD}
\rho\pard{\vek u}{t}+\rho(\vek{u\cdot\nabla})\vek u=-\vek \nabla p+
\vek{j \times B}+\rho\vek{g} \\
\nonumber
\pard{\vek B}{t}=\vek{\nabla\times}(\vek{u\times B})-
\vek{\nabla\times}(\eta\vek j)\\
\nonumber
\pard{U}{t}+\vek{\nabla\cdot S}=\rho\vek{u\cdot g}\ .
\end{eqnarray}
The set of equations~(\ref{eq:MHD}) is solved by means of Finite
Volume Method (FVM). For the numerical solution it is first
rewritten in its conservative form. 
The (local) state of magneto-fluid is then represented by the vector
of basic variables  
$\vek{\Psi}\equiv(\rho,\rho\vek{u},\vek{B},U)$, where $\rho$,
$\vek{u}$, $\vek{B}$, and $U$ are the plasma density, plasma
velocity, magnetic field strength and the total energy density, respectively. 
The energy flux $\vek S$  and auxiliary variables -- plasma pressure $p$
and current density $\vek j$ -- are defined by the formulae:
\begin{eqnarray}
\nonumber
\nabla\times\vek B=\mu_0\vek j\\
\nonumber
U=\frac{p}{\gamma-1}+\frac{1}{2}\rho u^2+\frac{B^2}{2\mu_0}\\
\nonumber \vek S=\left(U+p+\frac{B^2}{2\mu_0}\right)\vek u- \frac{(\vek
u\cdot\vek B)}{\mu_0}\vek B+\frac{\eta}{\mu_0}\vek j\times\vek B \ ,
\end{eqnarray}
and $\vek{g}$ is the gravity acceleration at the photospheric level.
Microphysical (kinetic) effects enter into the large-scale dynamics
by means of transport coefficients -- here via a (generalized) resistivity
$\eta$. In general, the role of non-ideal terms in the generalized Ohm's
law increases as the current density becomes more concentrated via
current sheet filamentation. To quantify this
intensification we use the current-carrier drift velocity
$v_{\rm D}=|\vek{j}|/(e n_e)$
as the threshold for non-ideal effects to take place. Such behavior
is presumed by theoretical considerations and confirmed by kinetic 
(Vlasov and PIC codes) numerical experiments 
\citep{Buchner+Elkina:2005, Buchner+Elkina:2006, Karlicky+Barta:2008}. 
In particular, we assume the following law for (generalized)
resistivity \citep[see also][]{Kliem+:2000}:
\begin{equation}
\label{eq:eta}
\eta(\vek r,t)=\left\{
  \begin{array}{lll}
    0 & : & |v_{\rm D}|\le v_{cr}\\
    C\frac{\left(v_{\rm D}(\vek r,t)-v_{cr}\right)}{v_0}& : &
    |v_{\rm D}|> v_{cr}
  \end{array}
  \right.
\end{equation}


\begin{figure}[t]
\epsscale{1.1}
\plotone{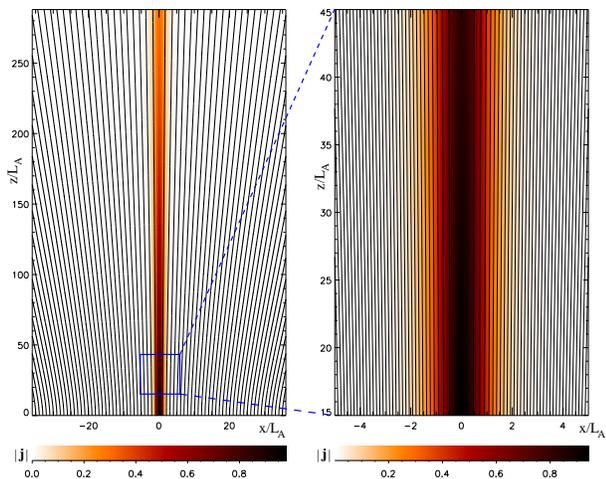}
\caption{Projection of the initial state ($t=0$) to the
  $xz$-plane. Black lines represent  the magnetic field lines, red
  areas the magnitude of the current density, its scale is given
  underneath. Enhanced view on selected area (the right panel) can be
  directly compared with the third panel in Fig.~\ref{fig:zoom1}.
}
\label{fig:init}
\end{figure}


In order to study the energy-transfer cascade it is appropriate to cover a
large range of scales. For structured grids it means the utilization of very 
fine meshes. For a given simulation box size the number of finite grid cells is
limited technically by CPU-time and memory demands. 
Alternatively, one can use a refined
mesh only at locations where the small-scale dynamics becomes important --
this idea forms the base for the Adaptive Mesh Refinement (AMR)
technique \citep[see, e.g.,][]{Berger+Oliger:1984,Dreher+Grauer:2005,
VanDerHolst+Keppens:2007}. 
We implemented the numerical solver for the MHD system of
equations~(\ref{eq:MHD}) in 
the form of block AMR Finite-Volume Method (FVM) code: 
Whenever the current sheet 
width drops below a certain threshold, a refined mesh sub-domain is created
and initialized with values one step backward in time. Its
evolution is then computed using accordingly refined time-step
(this procedure is commonly known as sub-cycling). The global dynamics
influences the sub-domain evolution 
by means of time- and spatially varying (interpolated) boundary
conditions. For the details of our AMR algorithm see \citet{Barta+:2010a}.
As an illustration, in Fig.~\ref{fig:zoom_cells} the regions of
enhanced grid resolution at  
$t=367.0 \tau_{\rm A}$ (for units used see below) are depicted at the
background of current density and magnetic field.
 

\begin{figure*}[t]
\plotone{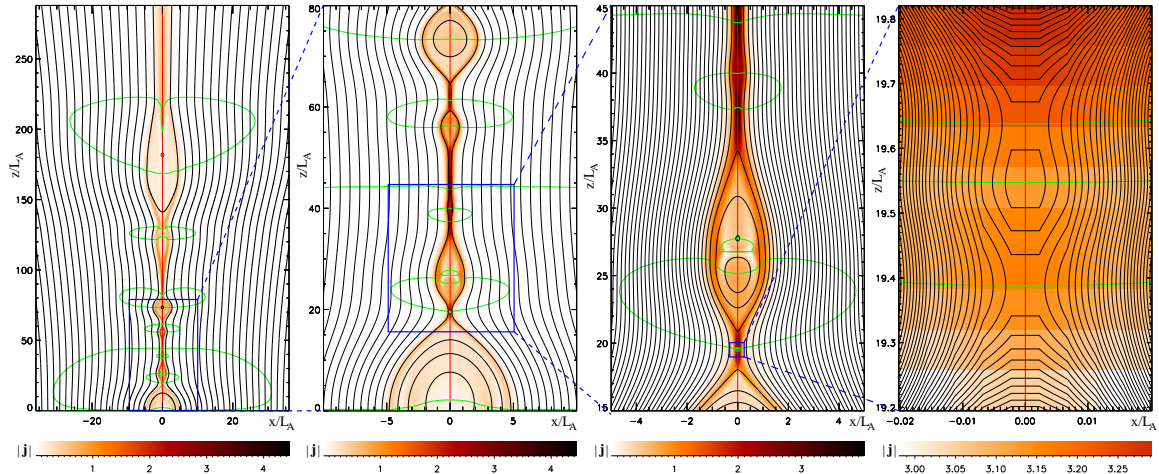}
\caption{Fragmentation of the flare current layer at $t=316$. Increasing zoom
  reveals further smaller magnetic structures (islands/plasmoids). 
  Red and green lines are positions of
  $B_x=0$, and $B_z=0$, respectively; their intersections represent
  the X- and O-type  ``null'' points.}
\label{fig:zoom1}
\end{figure*}


The partial differential equations~(\ref{eq:MHD}) are of a mixed
hyperbolic-parabolic type.
We utilized the time-splitting approach for their solution: First,
hyperbolic (conservative) part is solved using a second-order FVM
leap-frog scheme. In a second step the magnetic-diffusivity term is solved by
means of a (semi-implicit) Alternating-Direction-Implicit (ADI) scheme 
\citep{Chung:2002}. 

We solve the MHD system of Eqs.~(\ref{eq:MHD}) in a 2D
simulation box initially on a global (coarse) Cartesian grid.
The horizontal and vertical dimensions of calculated box are 800 and 
6400~grid cells, respectively. Using mirroring boundary condition at 
$x=0$  in the symmetric CS (see below)
we obtain a doubled box with an effective grid of $1600 \times 6400$ cells 
\citep[see also][for details]{Barta+:2008a}. 
We use
the following reference frame: The $z$-axis corresponds to the 
vertical direction, the $y$-axis is the
invariant (i.e. $\partial/\partial y=0$) direction along the PIL.
The $x$-axis is
perpendicular to the current layer and centered at the initial current
maximum. The
simulation is thus performed in the $xz$-plane, while the $xy$-plane 
corresponds to the solar photosphere; the PIL is located at $x=0$, $z=0$ 
\citep[see Fig.~5 in][]{Barta+:2008a}.

The simulation is performed in dimensionless variables. They are
obtained by the following normalization: The spatial
coordinates $x$, $y$, and $z$ are expressed in units of the current
sheet half-width $L_{\rm A}$ at the photospheric level ($z=0$).
Time is normalized to the Alfv\'en-wave transit time 
$\tau_{\rm A}=L_{\rm A}/V_{\rm A,0}$ through the current sheet, where 
$V_{\rm A,0}=B_0/\sqrt{\mu_0 \rho_0}$ is the asymptotic value
at $x\rightarrow\infty$ and $z=0$ of the Alfv\'en speed at $t=0$. 
Eq.~(\ref{eq:eta}) for anomalous resistivity in the dimension-less
variables then reads $\eta=C(|\vek{j}|/\rho-v_{cr})$ for
$|\vek{j}|/\rho > v_{cr}$. $C$ and $v_{cr}$ are now
dimension-less parameters. We used $C=0.003$ and $v_{cr}=15.0$ in our 
simulation.
The choice of the threshold $v_{cr}$ is not arbitrary as it 
is closely related to 
the numeric resolution reached by the code -- see the discussion in 
Section~\ref{sect:discussion}. The parameter $C$ was adjusted to reach peak
anomalous resistivities in the order of $10^6$ times higher than the Spitzer 
resistivity in the solar corona -- similar values for resistivity based on
non-linear wave-particle interaction are indicated by Vlasov simulations 
\citep{Buchner+Elkina:2006}.      

If not specified otherwise, all quantities in the paper
are expressed in this dimension-less system of units.
In order to apply our results to actual solar flares
appropriate scaling of dimension-less variables, however,
has to be performed. 
The gravity stratification included in our model introduces a natural 
length scale. Assuming an
ambient coronal temperature of $T=2$~MK the corresponding scale-height
for a fully-ionized hydrogen plasma is $L_{\rm G}=120$~Mm. The value
used in our simulation is $L_{\rm G}=200 L_{\rm A}$, hence $L_{\rm A}=600$~km.
For this scaling the flare arcade loop-top is
$\approx 10000$~km high, which corresponds well to observed
values. 
The initial CS width $2 L_{\rm A}=1200$~km
is roughly in line with the fact that the CS was formed by 
stretching of the magnetic field in the trail of ejected flux-rope/filament
which  itself has typical transversal dimensions $\approx 5000$~km 
\citep{Vrsnak+:2009}. It also corresponds (by order of magnitude) to
estimations made from observations of thin layers trailing behind CMEs that are
sometimes interpreted as signatures of current sheets 
\citep{Ko+:2003,Lin+:2007}.  
For the ambient magnetic field in the vicinity of the current layer we assume
 $B_{z0}=40$~Gauss \citep[see, e.g., the discussion in][]{Kliem+:2000}.

The initial state has been chosen in the form of a vertical
generalized Harris-type CS with the
magnetic field $\vek{B=\nabla\times A+\hat{e}_y} B_y$ slightly
decreasing with height $z$ \citep{Barta+:2010a}:
\begin{eqnarray}
\label{eq:init}
\nonumber
\vek{A}(x,y,z;t=0)=-B_{z0} \ln\left(e^\frac{x}{w_{\rm CS}(z)}+
e^{-\frac{x}{w_{\rm CS}(z)}}\right) \vek{\hat{e}_y} \\
B_y(x,y,z;t=0)=B_{y0}\\
\nonumber
\rho(x,y,z;t=0)=\rho_0\exp(-\frac{z}{L_{\rm G}})\ .
\end{eqnarray}
In the following we will refer to $B_x$ and $B_z$ as the
\textit{principal components} and $B_y$ as the \textit{guide field}.
The characteristic width of the initial current sheet varies with $z$ as
$$
w_{\rm CS}(z)=\frac{d\cdot z^2+z+z_0}{z+z_0}
$$
and $B_{y0}$, $B_{z0}$, $\rho_0$, $d$, and $z_0$ are appropriately chosen
constants: $B_{y0}=0.2$, $\rho_0=1.0$, $B_0=\sqrt{B^2_{y0}+B^2_{z0}}=1.0$, 
$d=0.003$, and $z_0=20.0$. 
The initial state given by Eq.~(\ref{eq:init}) corresponds
to a stratified atmosphere in the presence of gravity (which is
consistent with Eqs.~(\ref{eq:MHD})). 
The divergence of the magnetic field-lines towards
the upper corona is in agreement with the expansion of the coronal
field. It also favors up-ward motion of secondary plasmoids formed
in the course of CS tearing \citep{Barta+:2008b}. This leads to further
filamentation of the current sheets which develop between the
plasmoids \citep{Shibata+Tanuma:2001}. The current density and
magnetic field at the initial state are
displayed in Fig.~\ref{fig:init}. A rather thick current layer is
visible. An enhanced view of the selected area is
presented in the right panel for the sake of direct comparison with the 
current-density filamentation
developed in later stages of the evolution (see below in Fig.~\ref{fig:zoom1}).

Free boundary conditions  are applied to the upper and right part of 
the actually calculated right half of the box.
It means that von Neumann prescription $\partial/\partial\vek{n}=0$ has to be 
fulfilled
for all calculated quantities except the normal component of magnetic field 
$B_n$ and its contribution to the total energy density $U$. 
$B_n$ and $U$ are extended in the second step 
fulfilling $\vek{\nabla\cdot B}=0$.
In order to satisfy MHD boundary conditions symmetric ($q(-z)=q(z)$) relations
are used for $\rho$ 
$B_y$, $B_z$, $U$, and anti-symmetric ($q(-z)=-q(z)$) 
for $B_x$ at the bottom boundary while velocities 
are set to zero there ($\vek{u=0}$). This ensures that the
principal magnetic-field  
component is vertical at the bottom boundary and that the total magnetic
flux passing through that boundary does not 
change on the rather short time-scales of the eruption, as enforced 
by the presence of a dense solar photosphere \citep{Barta+:2008a}.
Mirroring boundary conditions (symmetric in $\rho$, $u_y$ 
$B_x$, $B_y$, and $U$  and antisymmetric in $u_x$ and $B_z$) 
are used for the left part of boundary at $x=0$ (=the center of the CS).
We use these symmetries to construct (mirror) the left half of the full
effective box.


\begin{figure*}[t]
\plotone{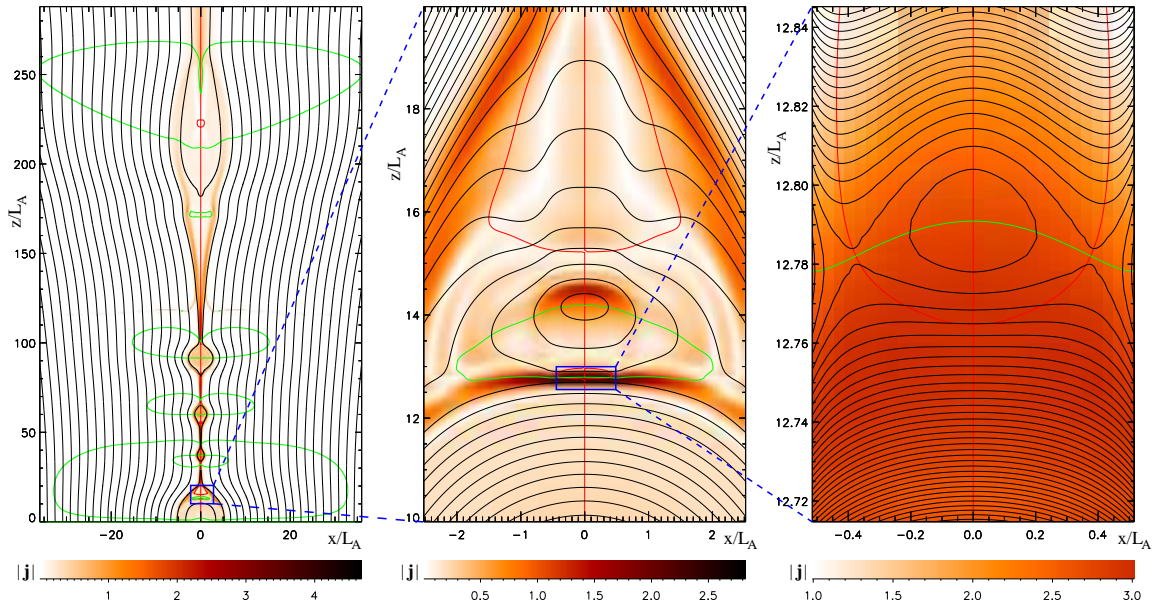}
\caption{Fragmentation of the current layer in the transversal
  direction at $t=367$. Enhanced zoom reveals further tearing 
  in the transversal
  (horizontal) current sheet formed due to the mutual interaction
  (merging/coalescence) between plasmoid and the loop-arcade. Red and green
  lines are as in Fig.~\ref{fig:zoom1}.}
\label{fig:zoom2}
\end{figure*}


The asymptotic plasma beta parameter at $x\rightarrow\infty$ and $z=0$ 
is $\beta=0.1$ and the ratio of specific heats is $\gamma=5/3$
(adiabatic response). 

The coarse-mesh sizes are $\Delta x=\Delta z=0.045$
in the dimensionless units. Thus, with the reference frame established above 
the entire box corresponds to
$\langle -36,36\rangle\times\langle 0,288\rangle$ in the 
$xz$-plane. The simulation was performed over 400 normalized Alfv\'en times.
To save disk space only the most interesting interval $t=300$ --- $400$
has been recorded with a step of $0.5\ \tau_{\rm A}$. 

Note that the initial state described by Eq.~(\ref{eq:init}) is not an
exact MHD equilibrium. Nevertheless, the resulting field variations are
much weaker than those introduced by reconnection.
 
At the very beginning, in order to trigger reconnection, the system is
perturbed by enhanced resistivity localized in a small region
surrounding a line $x=0$, $z=30$ in the invariant
direction $y$ for a short time $0\le t \le 10$ 
\citep[see also][]{Magara+:1996}. This short perturbation sets a localized
inflow which somewhat compresses the current layer around the selected
point. It should mimic the effect of various irregularities that can
be expected during the CS stretching in actual solar eruptions, 
see also \citet{Riley+:2007}. Later, the
resistivity is switched on only if the threshold according to
Equation~(\ref{eq:eta}) is exceeded. As the threshold for anomalous resistivity
cannot be reached for a coarse grid
(the threshold for mesh-refinement is reached earlier than the threshold for
anomalous resistivity onset),
the condition in
Eq.~(\ref{eq:eta}) is actually checked only at the smallest resolved
scale, for the large-scale dynamics we take $\eta=0$.


\section{Analysis of model results}
\label{sect:results}

We used the above described numerical code in order to study, which
mechanisms are involved in the transfer of free magnetic energy from
large to small scales. Thanks to the adaptive mesh we
were able to cover scales from $4.5\times 10^{-3} \ L_{\rm A}$ to
$\approx 300 \ L_{\rm A}$ (the larger size of the simulation box),
i.e. over almost five orders of magnitude.

The early system evolution can be briefly described as follows:
After the localized initial resistivity pulse a flow pattern sets-up 
that leads to CS stretching (in the $z$-) and compression 
(in the $x$-direction). 
Eventually, the condition in  Eq.~(\ref{eq:eta}) for anomalous resistivity 
is reached at the smallest resolved scale and first tearings occur. Dynamics
of the plasmoids formed by the tearing process leads to further stretching of
CSs interleaving the mutually separating plasmoids. This leads to further
generation of tearing.  
Later, after $t\approx 300$ the smallest magnetic 
structures yet consistent with the resolution start appearing.  
Here we present an analysis of this more developed stage of cascade.
Results are shown in Figs.~\ref{fig:zoom1}
and~\ref{fig:zoom2}. Fig.~\ref{fig:zoom1} shows the state of magnetic
field and current density at  $t=316$. For better
orientation auxiliary lines are added indicating the locations where $B_x=0$
and $B_z=0$. Their intersections show the positions of
O-type and X-type ``nulls'' -- the points, where only the guide field remains
finite. Areas indicated by blue line are consecutively zoomed (from
left to right panels). The left-most panel shows 
the entire simulation box and the right-most corresponds to a zoom at
the limit of the AMR-refined resolution. 
The figure shows, how parts of the current
layer are stressed and thinned between separating
magnetic islands/plasmoids formed by tearing instabilities. 
The current-layer filamentation is the most clearly pronounced if one compares 
the zoomed views of the same selected area at the initial state 
(Fig.~\ref{fig:init}, right panel) and the system state at $t=316$
(Fig.~\ref{fig:zoom1}, the third panel). During the dynamic evolution the
even thinner, stretched current layers become, after some time, unstable
to the next level of tearing and even smaller plasmoids are
formed. The zoomed figures show that cascading reconnection has formed
plasmoids at the smallest resolved scales: The $x$-sizes of
the largest and smallest resolved plasmoid in Fig.~\ref{fig:zoom1}
range from $\approx 10$ down to $\approx 0.01$,
the $z$-sizes are from $\approx 0.2$ to $\approx 70$. 

Plasmoids formed by tearing instability are not only
subjected to the separation but they can  also approach each
other. As a result the magnetic flux
piles-up and transversal (i.e. horizontal, perpendicular to the original
current layer) current sheets are formed between pairs of plasmoids approaching
each other. Earlier simulations with lower effective resolution treated
the plasmoid merging as a coalescence process without any
internal structure of the small-scale (sub-grid) current sheet between
the magnetic islands since their thickness was not resolved
\citep{Tajima+:1987, Kliem+:2000, Barta+:2008b}. If resolved, however,
the transversal current sheet does not just dissipate. Instead it is
subjected to the tearing instability in the direction perpendicular to
the primary current layer. This is shown in Fig.~\ref{fig:zoom2}. 
The most detailed resolution (right-most panel) clearly reveals
the formation of the O-point at 
$x=0 L_{\rm A}$, $z=12.79 L_{\rm A}$ and two adjacent X-points at 
$x=0.37 L_{\rm A}$, $z=12.78 L_{\rm A}$ and 
$x=-0.37 L_{\rm A}$, $z=12.78 L_{\rm A}$. 
We call this process ``fragmenting coalescence'' in order to emphasize that
even smaller structures are formed during the merging of two plasmoids.
Thus both the tearing and (fragmenting) coalescence processes contribute 
to the fragmentation of the original thick and smooth current layer.

\subsection{Fragmentation of CS: Scaling}
\label{subsect:scaling}


\begin{figure}[t]
\epsscale{1.0}
\plotone{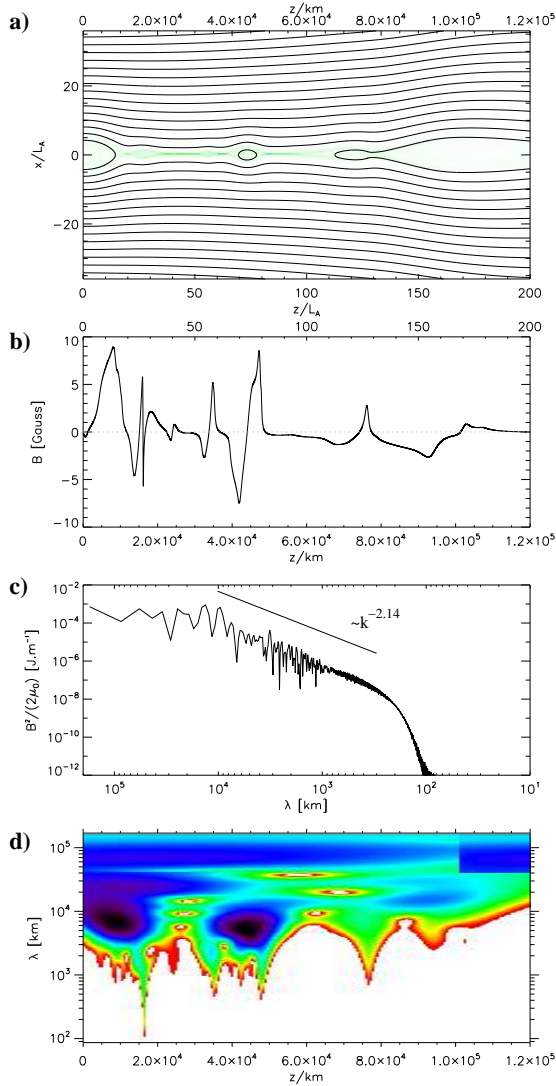}
\caption{The 1D scale analysis of the magnetic field structure along the
  line $x=0$. (a)
  Magnetic field lines and the current density structure (green) in the
  computational domain at $t=316$. The $z$-axis shows
  positions both
  in the units of $L_{\rm A}$ (top) and in kilometers according to
  scaling adopted in Section~\ref{sect:model}. (b) Profile of the $B_x$
  component of magnetic field along the line $x=0$. (c)
  Fourier power spectrum of the $B_x$ profile. (d) Wavelet power
  spectrum  of the $B_x$ profile.}
\label{fig:scaling}
\end{figure}


\begin{figure}[t]
\epsscale{1.1}
\plotone{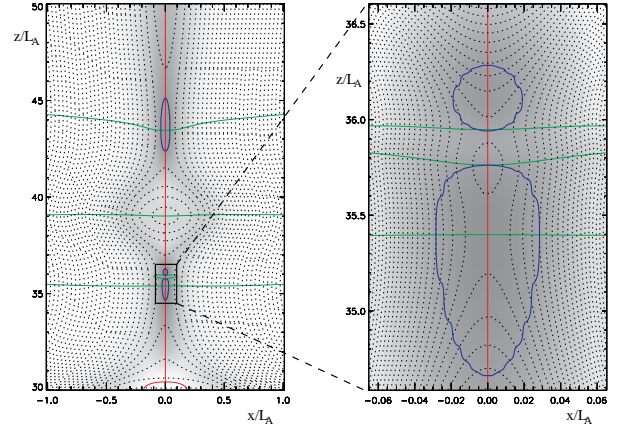}
\caption{Filamentation and splitting of diffusive regions at $t=328$. 
  Detailed treatment shows how the magnetic diffusivity is concentrated into
  multiple thin channels. The blue-line bounded areas indicate the
  diffusion regions where the 
  generalized anomalous resistivity (Eq.~\ref{eq:eta}) is
  finite. Field lines are dashed to be better distinguished from the
  resistive region boundaries. Current density is represented by gray levels.
  The right panel displays detailed view of the
  rectangular box in left. Red and green
  lines are as in Fig.~\ref{fig:zoom1}.}
\label{fig:dissip}
\end{figure}


In order to study scaling properties of the continued fragmentation
of the magnetic structures associated with the current layer we
performed both a 1D Fourier and a wavelet analysis of the magnetic field along
the vertical axis $\{[x=0,y=0,z \in\langle 0, 288 \rangle ]\}$. 
We use for this study the $B_x$
component since $B_z=0$ there due to the boundary condition. The results
are shown in Fig.~\ref{fig:scaling}. The upper panel shows the
magnetic field and current density in the sub-set of the entire
computation domain ($z\in\langle 0, 200\rangle$, note the rotated view),
where the current layer is fragmented. Panel (b) shows the profile of $B_x$
along the current-layer axis, and panels (c) and (d) the Fourier and
wavelet analyses of this profile. The Fourier power spectrum exhibits
a power-law scaling with the spectral index $s=-2.14$ 
in rather wide range of scales 300~km -- 10000~km. 
This clearly indicates cascading nature of the continued fragmentation. 

The energy-transfer cascade ends at $\approx 300$~km in 
Fig.~\ref{fig:scaling}. 
This is closely related to the dissipation
threshold $v_{cr}$ that has been chosen as $v_{cr}=15.0$ in our simulation. By
selecting this value we shifted dissipation-scale domain into
the window of resolved scales -- see the discussion in 
Section~\ref{sect:discussion}. Typical width of dissipative current
sheets in our model is thus $L_d\approx L_A/15.0=40$~km. Since the
plasmoid dimensions along the CS are about one order of magnitude 
(typically 6$\times$) larger than across CS, distribution of magnetic energy
in structures along CS, which is depicted in Fig.~\ref{fig:scaling},
reaches its dissipation scale at $\approx 6\times 40\ {\rm km}=240$~km.
In reality, the ion inertial
length $d_i=c/\omega_{pi}$ is considered as a typical width 
of dissipative current sheets \citep{Buchner:2007}. 
Its value for parameters $B_0$ and $V_{\rm A}$ used
in this paper is $d_i\approx 4$~m in the simulation-box center,
i.e. at $z\approx140$ (see Fig.~\ref{fig:transfer}).

The most pronounced features in the wavelet
spectrum are the locations of low signal (the white islands). They
correspond to the filamented parts of the current layer between
plasmoids. Their distribution indicates, that the filamented current
sheets are embedded within the global current layer in a hierarchical
(qualitatively self-similar) manner.

\subsection{Fragmentation of CS: Diffusion regions}
\label{subsect:acceleration}

The current sheets are filamented down to the resolution limit of
our simulations (in reality to the kinetic scales). The smallest
current-density structures contain dissipative/acceleration regions. 
In the following we will study the 
structure, distribution and dynamics of these non-ideal
regions embedded in the global current layer.

Cascading reconnection and consequent fragmentation of the current
layer may have significant impact
also for particle acceleration in solar flares. 
Instead of a single diffusion region assumed in the 'classical' picture
of the solar reconnection, cascading fragmentation causes the formation of
large amount of thin non-ideal channels. 
The structuring of non-ideal regions in our simulation is 
depicted in Fig.~\ref{fig:dissip}.
The left panel shows two areas of dissipation around 
$x=0$, $z=36$ and $x=0$, $z=44$. A closer look (right panel, 
note the large zoom), however, reveals that the bottom
dissipation region is structured and it is in fact formed by two
regions of finite magnetic diffusivity that are associated with two
X-points at $x=0$, $z=35.40$ and $x=0$, $z=35.97$ interleaved with a (micro)
plasmoid.
The multiple
dissipative regions embedded in the global current layer are
favorable for efficient (and possibly multi-step) 
particle acceleration. At the same time they
provide a natural explanation of \textit{fragmented energy release} as
it has been inferred from HXR and radio observations
\citep{Aschwanden:2002, Karlicky+:2000}. Since they are  embedded in
the large-scale current layer the 'classical' well-organized global
picture of eruptions is kept simultaneously.

Fig.~\ref{fig:dissip} shows that 
the X-points formed in the thinned current sheets between magnetic islands
are connected with the
thin channels of magnetic diffusivity. Hence it is appropriate to study the
distribution and dynamics of these non-ideal regions by means of
tracking the X-points associated with them. We present such analysis in 
Fig.~\ref{fig:nulls}. 
In order to see the ``skeleton'' of the reconnection dynamics we followed
the positions of all magnetic ``null'' points during the entire
recorded interval $t=300$ --- $400$. 
The results show
the kinematics of the O-type (red circles) and X-type points. 
Motivated by our endeavour to establish a closer relation of the model
to observable quantities in our consecutive 
study \citep{Barta+:2010c} we also paid special attention to the magnetic 
connectivity of the X-points to the
bottom boundary. For this sake, the X-points connected to the model base
(= the photosphere) are painted as green asterisks while the unconnected
X-points are displayed as black crosses.

Since we are interested in the
cascade that already developed into the stage reaching the smallest resolved
structures ($t\ge300$), many X- and O- points connected with the larger-scale 
(and therefore longer-living) structures (plasmoids) are already formed and 
their space-time trajectories enter Fig.~\ref{fig:nulls}(a) from the bottom. 
Fig.~\ref{fig:nulls} thus shows mainly full life-cycles of the X- and O- points 
related to the smallest resolved plasmoids. 
It is best visible in the three bottom panels (b) -- (d) that show
zoomed views (projected to the $zt$-plane) of typical 
examples of null-point dynamics -- the creation of temporary X--O pairs (panels
(b) and (d)) and plasmoid merging (c). As it can be seen from 
panels (b) and (d) the X-points can become magnetically connected (the
right X-point in panel (d)) or disconnected (panel (b)) to/from
the bottom boundary during their lifetime. Note also the
splitting (and subsequent merging) of the X-point at $x=0$, $z=12.8$
into an X-O-X configuration  between $t\approx 360$ 
and  $t\approx 380$ in panel (a). This process maps
the tearing in the transversal (horizontal) current sheet formed
between interacting plasmoid and the loop-arcade (see also 
Fig.~\ref{fig:zoom2}).
Note that Fig.~\ref{fig:nulls} can be compared with Fig.~5 in
\citet{Samtaney+:2009}. The main difference is just in the presence of
the off-plane X-points formed by the fragmentation of the CS between
coalescing plasmoids in our simulation.


\section{Discussion}
\label{sect:discussion}

Reconnection in the trailing current layer behind an ejected flux-rope
(filament) is a key feature of the 'standard' CSHKP scenario of solar flares.
A large amount of free magnetic energy is accumulated around this rather thick
(relative to plasma kinetic scales)
and very long layer. The thickness of this layer was estimated both from 
the observed brightening \citep{Ko+:2003, Lin+:2007} and based on a typical 
transversal dimensions of a filament \citep{Vrsnak+:2009}.
Both ways one obtains the order of magnitude of $\approx 1000$~km. 
On the other hand collisionless 
reconnection requires dissipation at very small scale,
thin current sheets with typical width of the order of
$\approx 10$~m in the solar corona \citep{Buchner:2007}. 
The fundamental question arises how the accumulated 
energy is transferred from large to small scales. 
Or, in other words, what are the mechanisms of direct energy cascade in
magnetic reconnection.
We addressed this question using high-resolution AMR 
simulation covering broad range of scales to investigate 
the MHD dynamics of an expanding current layer in the solar 
corona.

\subsection{Mechanisms of direct energy cascade}

Our simulations reveal the importance of a continued
fragmentation of the current layer due to the interaction of two 
basic processes:
The tearing instability of stretched current sheets and the 
fragmenting coalescence of flux-ropes/plasmoids formed by the tearing
and subsequently forced to merge by the tension of
ambient magnetic field. 
After ejection of the primary flux-rope (i.e. the filament/CME),
a trailing current layer is formed behind it which becomes long and 
thins down. As it has been pointed out by theoretical analysis 
by \citet{Loureiro+:2007}, current layers with high enough length-to-width 
ratio become unstable for fast plasmoid instability. Moreover,
any irregularity in the plasma inflow that stretches the sheet facilitates
the tearing \citep{Lazarian+Vishniac:1999}.

Plasmoids that are formed are subjected to the tension of 
ambient magnetic
field, which causes them to move \citep{Barta+:2008b}. 
The motion can lead to their increasing separation. 
A secondary current layer then formed between them becomes,
again, stretched and a secondary tearing instability can 
take place. 
This simulation result, illustrated by Fig.~\ref{fig:zoom1}, fully confirms
the scenario suggested by \citet{Shibata+Tanuma:2001}, 
developed further by \citet{Loureiro+:2007} and
\citet{Uzdensky+:2010} into the analytical theory of chain plasmoid 
instability. The results are also in qualitative agreement 
with the simulations of plasmoid instability by \citet{Samtaney+:2009} and 
\citet{Bhattacharjee+:2009}, which has been performed, however, with
constant resistivity.


\begin{figure*}[t]
\epsscale{1.0}
\plotone{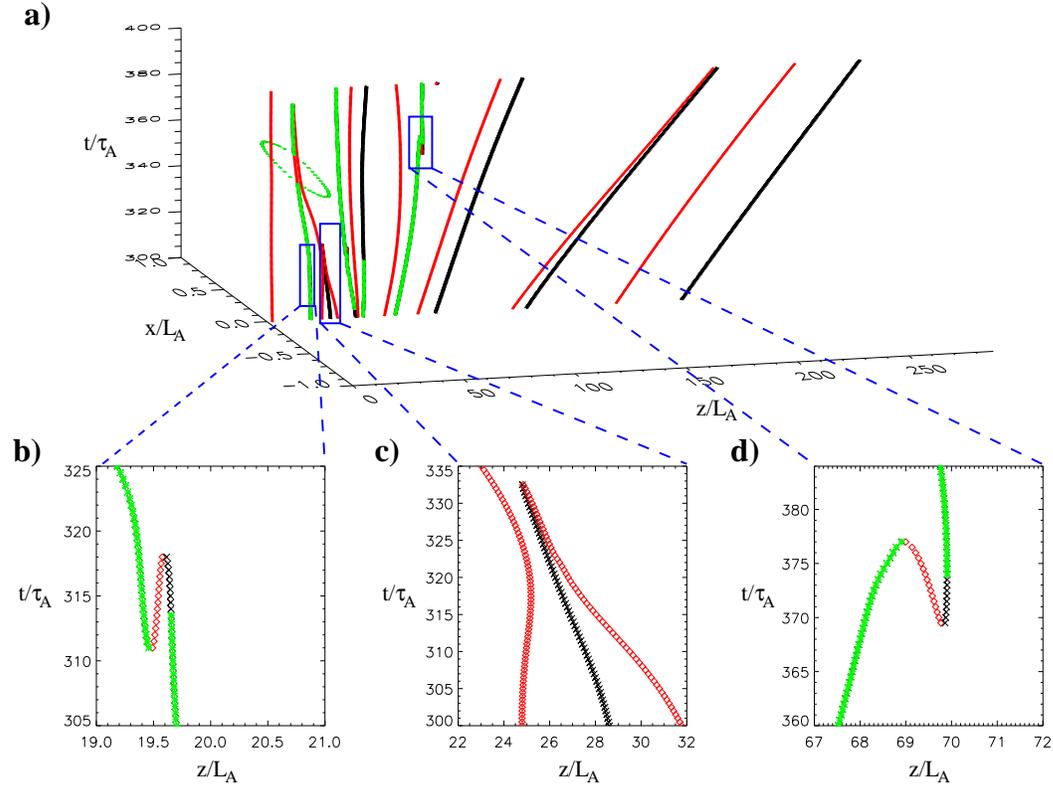}
\caption{Kinematics of the magnetic ``null'' points
  (X-points/dissipative regions and O-points/plasmoids). Red circles
  denote O-point positions, black crosses the X-point positions and
  green asterisks show the positions of those X-points which are
  magnetically connected to the photosphere. Three bottom panels show
  detailed view of selected rectangles and represent typical processes
  of X- and O-point dynamics: Bifurcation (null-pair creation) and merging
  (null-pair annihilation).}
\label{fig:nulls}
\end{figure*}



\begin{figure}[t]
\epsscale{1.1}
\plotone{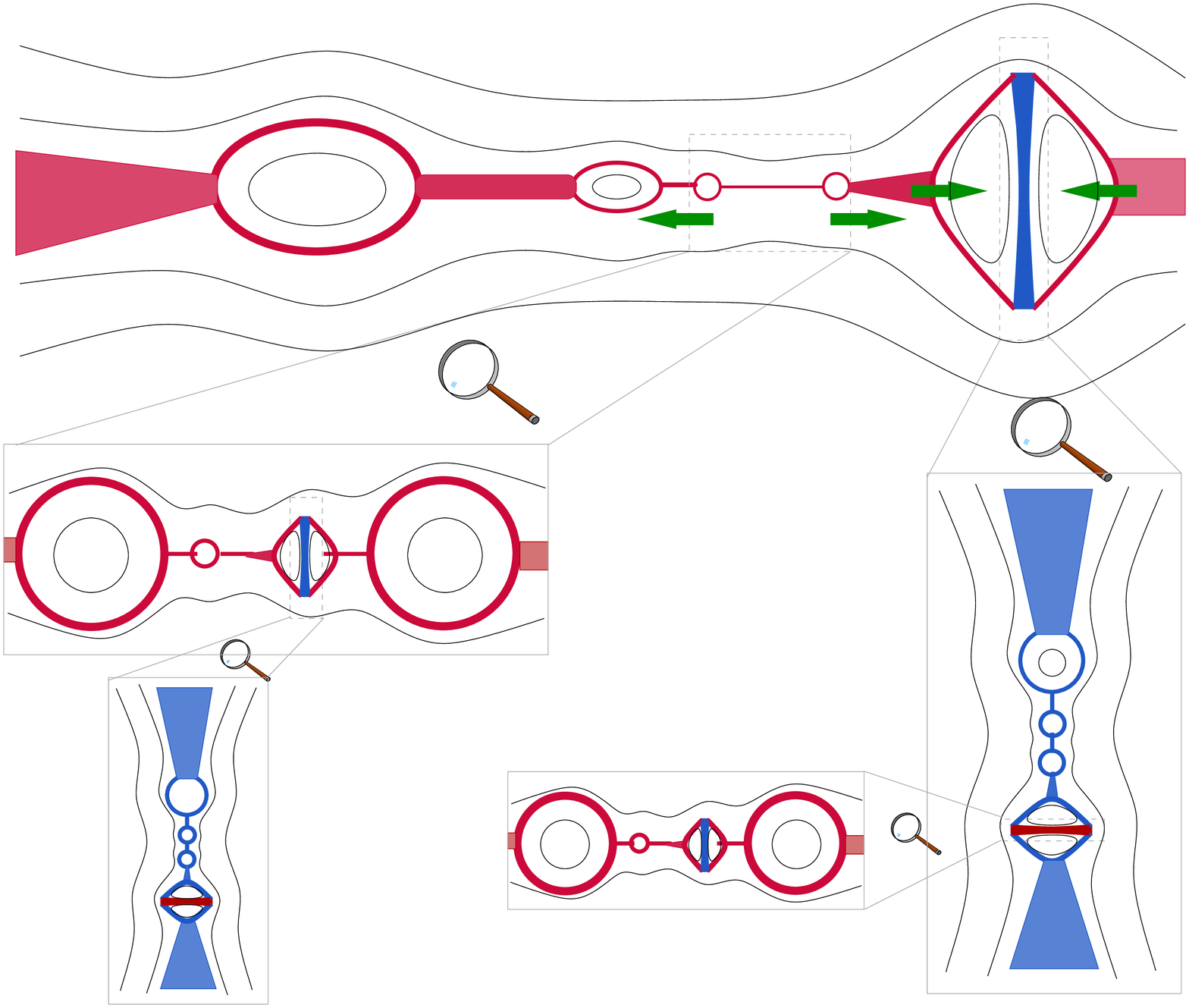}
\caption{Schematic view of cascade of fragmentation of the current
  layer by tearing and driven coalescence processes. Increasing zoom
  shows similar kind of processes repeating on smaller scales. Red and
  blue areas of various hue indicate various intensities of positive and
  negative out-of-plane ($j_y$) component of current density.} 
\label{fig:fragmentation}
\end{figure}


In addition to that, our simulation has shown that the converging motion 
of plasmoids leads to a magnetic-flux pile-up between mutually
approaching plasmoids. Consequently, secondary (oppositely directed) 
current sheets are formed
perpendicular to the original current layer. 
While previous studies found only unstructured current 
density pile-ups between merging magnetic islands 
our enhanced-by-AMR resolution reveals secondary tearing mode 
instabilities that take place in the transversal to the 
primary current sheet direction (see Fig.~\ref{fig:zoom2}).
This process represents a new mechanism of fragmentation and changes our view
to coalescence instability, which has been hitherto commonly considered as a
simple merging process of two plasmoids 
contributing to the inverse energy cascade only.
Note that this behavior is different from that seen for plasmoids 
at the dissipation scales in PIC simulations
\citep{Drake+:2005, Karlicky+Barta:2007}, where plasmoids merge
without subsequent tearing.

One can suppose that with even higher spatial resolution 
one would see more subsequent tearing mode instabilities 
altering with the fragmenting coalescence of the resulting magnetic 
islands/flux-ropes. As a result third- and higher order
current sheets could form. 

To sum up, the results of our simulation support the idea
that both the tearing \citep{Shibata+Tanuma:2001, Loureiro+:2007,
  Uzdensky+:2010} and ``fragmenting coalescence'' processes
lead to the formation of consecutively smaller magnetic structures
(plasmoids/flux-ropes) and associated current filaments. 
Subsequent stretching and compression cause a filamentation
of the current. This situation is schematically depicted in 
Fig.~\ref{fig:fragmentation} which can be seen as a generalization of
the scheme in Fig.~6. in \citet{Shibata+Tanuma:2001}.
One can expect that this cascade will continue down to the 
scales where the magnetic energy is, finally, dissipated.
Note that the physics and the corresponding scaling laws 
may change at intermediate (but still relatively small) 
scales when additional contributions to the generalized 
Ohm's law become significant, e.g. a Hall term 
-- see recent simulations by \citet{Shepherd+Cassak:2010} and
  \citet{Huang+:2010}.


\begin{figure}[t]
\epsscale{1.1}
\plotone{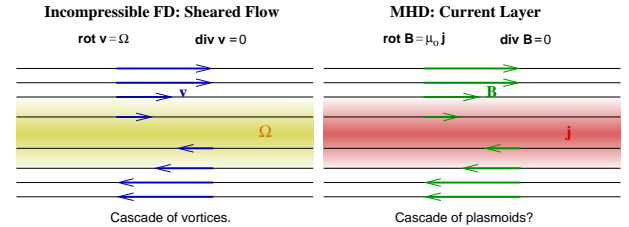}
\caption{Illustration of analogy between sheared-flow in
  incompressible fluid dynamics (FD) 
and magnetic reconnection in a large-scale current layers. In the case of
FD the mechanism of energy transfer from macroscopic to dissipative
(molecular) scale is known -- it is the cascade of vortex-tubes. The cascade of
magnetic flux-ropes/plasmoids can play this role in the case of magnetic
dissipation.} 
\label{fig:analogy}
\end{figure}


\subsection{Impact on reconnection efficiency}

Reconnection in the current sheet between merging plasmoids is fast
since it is driven by ambient-field magnetic tension which naturally pushes the
flux-ropes together.
Thus even shorter time-scales can be reached by this process than by 
tearing cascade in the stretched CS \citep{Shibata+Tanuma:2001}.  
And yet another point makes the overall reconnection 
process more efficient: Many magnetic-flux elements 
-- except those ejected out-wards to the escaping CME 
-- reconnect several times. 
First, during the primary tearing and plasmoid
formation and then again during plasmoid coalescence. 
Since coalescence leads to a follow-up tearing instability 
(in the transversal direction) the remaining magnetic flux 
is subjected to another act of magnetic reconnection.
This process resembles the \textit{recurrent 
separator reconnection} simulated by \citet{Parnell+:2008}.

\subsection{Relation to turbulence onset}

To some extent the initial situation of global, smooth and 
relatively thick sheets is similar to the turbulence on-set in
a sheared flow in (incompressible) fluid dynamics (FD) as
schematically shown in Fig.~\ref{fig:analogy}. 
Usually the typical length-scale of  shear flows --
the counterpart of the width of current layers -- 
is much larger than the dissipative (molecular) scale. 
The mechanism of energy transfer from large to small 
scales in classical FD is 
mediated by a cascade of vortex tubes: 
Large-scale vortices formed by shear flows can mutually 
interact giving rise to increased velocity shear at the
smaller scales in the space between them.
Each small shear flow element formed by this process can be, again, subjected
to this fragmentation. 
Based on our simulation results, we suggest a similar 
scenario for current-layer fragmentation. 
The role of the vortex-tubes in FD is in MHD taken over 
by flux-ropes/plasmoids. 

In analogy with the on-set of turbulence in sheared flows, one 
could expect that a dynamical balance would arise between fragmentation and
coalescence processes in later more developed stage. 
This should be manifested  by a power-law
scaling rule. 
Using 	AMR we reached a rather broad  (five orders of
magnitude) range of scales. This allowed us to perform 
a 1D scaling analysis of the magnetic-field structures 
formed along the current layer for the first time.
The scaling rule found exhibits, indeed, a power-law distribution with
the index $s=-2.14$ (Fig.~\ref{fig:scaling}).
Since our resolution still does not allow to make this scaling analysis only 
within the small selected subdomain around the CS center where one could 
expect isotropic 'turbulence' (we would lack sufficient range of scales for
that) it is difficult to compare the spectral index found over the whole
(clearly anisotropic) simulation domain with the values expected
from the theory for fully developed isotropic turbulence.

Obtained power-law distribution is also in qualitative agreement with the
concept of fractal reconnection by \citet{Shibata+Tanuma:2001} and
with hierarchical analytical model of plasmoid instability 
\citep{Loureiro+:2007}
as described by \citet{Uzdensky+:2010}. They use distribution
functions for plasmoid width and contained flux in order to characterize
statistical properties of plasmoid hierarchy rather than power
spectrum. We plan to perform similar analysis of our results in the
future study in order to compare the results also quantitatively.

In order to obtain as broad as possible scale range in the plane where
reconnection occurs, we performed these simulations using 2.5D
approach. The question arises to what extent a full-3D treatment would
change the resulting picture. 
In the FD cascade vortices are deformed, their cross-sections 
change along their main axis, even in the topological sense. 
The object defined as a single vortex tube in one place can be split 
into two in another location. 
One can expect a similar behavior of plasmoids/flux-tubes
in MHD. There they could be subjected to the kink and similar
instabilities with $k_y>0$.  
Such processes would naturally lead to the modulation of reconnection
rate along the PIL. Observations indicating such effect have
already been presented \citep{McKenzie+Savage:2009}. 
To some extent the expected behavior can also be obtained
for kink instabilities of tiny current channels at the dissipative
scale studied by 3D PIC simulations 
\citep{Zhu+Winglee:1996, Karlicky+Barta:2008}.
Nevertheless, \citet{Edmondson+:2010} found no such evidence in their 3D AMR
MHD simulation, which uses, however, a different set-up. Sizes of the plasmoids
formed under 3D perturbation in the direction that corresponds to the 
invariant $y$-axis in our 2.5D case have been found very short preventing a
kink-like instabilities to develop. On the other hand, the resulting 
plasmoid lengths might depend on the initial guide field 
\citep{Edmondson+:2010,Dahlburg+:2005}.

To sum up, the answer to this question can be found only via full 
3D simulations with similar initial set-up as we used here. Therefore
we plan to extend our current 2.5D simulations 
with very high in-plane resolution with moderately resolved 
structuring in the third dimension.

\subsection{Fragmented energy release and particle acceleration}

Cascading fragmentation of the current layer is closely related to 
another puzzling question of current solar flare research -- the
apparent contradiction between observed regular large-scale dynamics
and signatures of fragmented energy release in (eruptive) flares. This
duality is reflected by two classes of flare models: The
'classical' CSHKP scenario based on magnetic reconnection in a
single global flare current sheet and the class of ``self-organized
criticality'' (SOC) models based on the avalanche of small-scale
reconnection events
in multiple current sheets formed as a consequence of either chaotic
\citep{Aschwanden:2002, Vlahos:2007} or regular but still complex 
boundary motions causing, e.g., magnetic braiding 
\citep{Wilmot-Smith+:2010}. 

The model of cascading reconnection has the potential 
to provide a unified view on these seemingly
very different (see the discussion in the next paragraph)
approaches. From the global point of view, it
coincides with the classical CSHKP model keeping the 
regular picture of the process at large scales.
At the same time, due to the \textit{internal} current-layer 
fragmentation the tearing/coalescence cascade forms 
multiple small-scale current sheets and potential 
diffusive regions. As a consequence, fragmented energy release, 
e.g., by particle acceleration, can take place in these tiny regions.
To some extent this finding can be seen as a follow-up of the 'bursty'
reconnection regime found by \citet{Kliem+:2000}. These authors show that 
intermittent signal (X-ray, radio) can be related to chaotic pulses of the
(resistive) electric field in the dissipation region around X-point. The
pulsed regime is a consequence of non-linear interplay between governing MHD
equations and the anomalous resistivity model. In our view, however, these
pulses result from the subgrid physics unresolved in earlier simulations.
Essentially, what has been seen as a single dissipative region around a single
X-point in the coarse-grid models is in fact a (``fractal-like'') set of 
non-ideal areas around multiple X-points (see Fig.~\ref{fig:dissip}) 
that interleave very small-scale mutually interacting plasmoids. 
This view based on high-resolved simulation is perhaps closer to the term of
``fragmented energy release'' that assumes the energy dissipation to be
performed via many concurrent small-scale events appearing in multiple 
sites distributed in space.
In this context it is interesting to note how surprisingly well
the phenomenological resistivity model (Eq.~\ref{eq:eta}) used by
\citet{Kliem+:2000} mimics the
sub-grid scale physics as it has been able to reproduce qualitatively temporal 
behavior of resistive electric field even without resolving actual processes
that are responsible for it.
Note also that a possible role of tearing and coalescence in 
fragmentation of the energy release in solar flares has been 
mentioned already by \citet{Kliem:1990}.

We would like to emphasize that there is a fundamental difference 
between the fragmented 
energy release by cascading reconnection and SOC models. It is
rooted in the fact that in cascading reconnection 
the complexity/chaoticity is due to an intrinsic current-layer 
dynamics, i.e., due to spontaneous fragmentation, while it is
introduced in SOC models through (external) boundary conditions (chaotic
boundary motions).
In fact, these two concepts are contrary to some extent: While in SOC-based
models the global flare picture is built as an avalanche of many small-scale
events (bottom$\rightarrow$top process in the scale hierarchy) in cascading
reconnection 
small scale structures are formed as a consequence of internal dynamics of
large-scale CS (top$\rightarrow$bottom process).

Fragmented energy release is closely related to the number
problem of particles accelerated in solar flares. 
A single diffusion region
assumed in the CSHKP model provides a far too small volume for accelerating
strong fluxes of particles as they are inferred from the HXR
observations. 
This argument has been used in favor of SOC-based models as they 
provide energetic-particle spectra and time-profile distributions
as observed and explain large energetic particle  fluxes. 

We suggest that, however, the inclusion of cascading reconnection
into the CSHKP has even more capabilities than SOC models.
It could explain both the distribution and the number of
accelerated particles, based on a physical consideration
of many small-scale current sheets which can host tiny
diffusive channels that all can act as the acceleration regions 
(see Fig.~\ref{fig:dissip}). 

Here it is appropriate to make one technical comment:
In an MHD simulation with resistivity model described by
Eq.~(\ref{eq:eta}), respectively its dimension-less version, the size
and the number of diffusive regions are controlled mostly by the threshold
$v_{cr}$ for the onset of (anomalous) diffusivity. 
The higher $v_{cr}$ is chosen, the thinner the current sheets 
can become, the smaller and more numerous are the 
embedded diffusion regions. 
Since one has to resolve these diffusive regions in the simulation,
one has to choose the threshold $v_{cr}$ low enough to be able
to resolve the dissipation regions appropriately.
In ideally resolved simulations, covering all scales down 
to the real physical dissipation length, the critical 
velocity $v_{cr}$ could be chosen of the order of physically
relevant value -- the electron thermal speed $v_{Te}$ 
\citep{Buchner:2007}. In dimensionless units this corresponds to 
$v_{cr}=L_{\rm A}/d_i \sqrt{m_i/m_e}\sqrt{\beta/2}$,
where $m_i$ and $m_e$ are the proton and electron masses.
For a technically limited spatial resolution, one has to choose a 
(much) smaller value of $v_{cr}$ in order to resolve the smallest
possible current sheets, before dissipation sets in, by a reasonable 
number of grid points. 
Since the resolution in our current AMR simulation is higher 
than in earlier models, we could choose a more reasonable value of 
$v_{cr}=15.0$ \citep[while older simulations used  $v_{cr}=3.0$ -- see, e.g.,]
[]{Kliem+:2000, Barta+:2008a}. This allowed us to track down
more fragmented, smaller reconnection regions. 
If we extrapolate this trend, we can expect that with even
higher resolution one would find even more and tinier diffusive regions.
They would be grouped hierarchically (self-similarly), 
occupying a sub-space of the global current layer. 
Such kind of distribution is indicated in the wavelet spectra (white
islands in Fig.~\ref{fig:scaling}(d)), and also by the positions and
motion of the associated X-points in Fig.~\ref{fig:nulls}.
The latter shows a structured grouping of ``null points'' and 
their various life times.

\section{Conclusions}

Our simulation has shown that cascading reconnection 
due to the formation and fragmenting coalescence of plasmoids/flux-ropes 
is a viable physical model of fragmented magnetic energy release 
in large-scale systems, like solar flares.
Cascading reconnection addresses at once three key
problems of the current solar-flare research: The scale-gap between 
energy-accumulation and dissipation scales, the duality between 
regular global-scale dynamics and fragmented energy-release signatures
observed simultaneously in solar flares, and the issue of particle
acceleration. All these problems arising from observations appear to
be tightly related via cascading reconnection.

In order to evaluate relevance of the cascading reconnection 
for actual solar flares further it is desirable, however,
to identify and predict model-specific observables and 
to search for them in observed data. 
We are going to propose possible specific signatures and 
compare them with observations in a consecutive 
paper \citep{Barta+:2010c}.


\acknowledgments 
This research was performed under the support of the
European Commission through the SOLAIRE Network (MTRN-CT-2006-035484)
and the grant P209/10/1680 of the Grant Agency of the
Czech Republic, by the grant 300030701 of the Grant Agency of the 
Czech Academy of Science and the research project AV0Z10030501 of 
Astronomical Institute of the Czech Academy of Science. The authors
thank to Dr.~Antonius Otto for inspirational discussions and to unknown
referee for valuable comments that helped to improve the quality of the paper.



\end{document}